\documentclass[orivec]{llncs}

\usepackage[utf8]{inputenc}
\usepackage[T1]{fontenc}

\usepackage{amsmath,amssymb}
\usepackage{graphicx}
\usepackage{xcolor}
\usepackage{array}
\usepackage{hyperref}
\usepackage{subcaption}
\usepackage{tikz}
\usepackage{fancyvrb}
\usepackage{pifont}
\usepackage{xspace}
\usepackage{rotating}
\usepackage{booktabs}
\usepackage{microtype}

\usepackage{todonotes}
\LetLtxMacro{\oldTodo}{\todo}
\renewcommand{\todo}[1]{\oldTodo{TODO: #1}}

\usepackage[normalem]{ulem}
\usepackage{MnSymbol}
\usepackage[autolanguage]{numprint}
\usepackage{etoolbox}

\usepackage{multirow}
\usepackage[inline,shortlabels]{enumitem}
\usetikzlibrary{shapes}
\usetikzlibrary{plotmarks}
\usepackage{tabto}
\usepackage{listings}
\usepackage{colortbl}
\usepackage{lipsum}

\newtoggle{anonymizesystems}

\newcommand{\SQLPGQ}{SQL\slash PGQ\xspace}
\newcommand{\gcore}{\mbox{G-CORE}\xspace}

\newcommand{\yedscale}{0.53}

\newcommand{\interactivevone}{Interactive~v1\xspace}
\newcommand{\interactivevtwo}{Interactive~v2\xspace}

\newcommand{\snbinteractivevone}{SNB Interactive~v1\xspace}

\newcommand{\eg}{e.g.\xspace}
\newcommand{\ie}{i.e.\xspace}

\newcommand{\etc}{etc.\xspace}
\newcommand{\vs}{vs.\xspace}

\definecolor{ldbcpale}{HTML}{86cd7d}
\definecolor{ldbc}{HTML}{439539}
\definecolor{red}{rgb}{0.7, 0.11, 0.11}
\definecolor{blue}{rgb}{0.0, 0.0, 0.55}
\definecolor{green}{rgb}{0.0, 0.42, 0.24}
\definecolor{grey}{rgb}{0.52, 0.52, 0.51}

\newcommand*{\thead}[1]{\multicolumn{1}{c}{\bf #1}}
\newcommand*{\theadbar}[1]{\multicolumn{1}{c|}{\bf #1}}
\newcommand*{\theadside}[1]{\multicolumn{1}{c}{\bf \begin{sideways}#1$\,$\end{sideways}}}
\newcommand*{\theadsidebar}[1]{\multicolumn{1}{c|}{\bf \begin{sideways}#1$\,$\end{sideways}}}

\newcommand*{\theadsideleft}[2]{\multicolumn{1}{l}{\bf \begin{sideways}#1$\,$\end{sideways} \begin{sideways}#2$\,$\end{sideways}}}
\newcommand*{\theadsidebarleftsingle}[1]{\multicolumn{1}{c|}{\bf \begin{sideways}#1$\,$\end{sideways}}}
\newcommand*{\theadsidebarleftdouble}[2]{\multicolumn{1}{l|}{\bf \begin{sideways}#1$\,$\end{sideways} \begin{sideways}#2$\,$\end{sideways}}}

\newcommand{\type}[1]{\textsf{#1}}

\newcommand{\Messages}{\type{Messages}\xspace}

\newcommand{\Persons}{\type{Persons}\xspace}

\newcommand{\yes}{$\bigotimes$}
\newcommand{\maybe}{$\bigoslash$}
\newcommand{\no}{$\bigcircle$}
\newcommand{\myast}{$\bigostar$}

\newcommand{\audited}{^\textrm{(a)}}
\newcommand{\stdestablishing}{_\textrm{(s)}}
\newcommand{\information}{^\textrm{(i)}}

\newcommand{\OneK}{\numprint{1000}\xspace}

\newcommand{\TenK}{\numprint{10000}\xspace}
\newcommand{\ThirtyK}{\numprint{30000}\xspace}
\newcommand{\HundredK}{\numprint{100000}\xspace}

\renewcommand{\paragraph}[1]{\noindent\uline{\textbf{#1.}}\xspace}

\definecolor{Person}{HTML}{fdb462}
\definecolor{Message}{HTML}{bebada}
\definecolor{Forum}{HTML}{b3de69}
\definecolor{Comment}{HTML}{80b1d3}
\definecolor{Post}{HTML}{fb8072}
\definecolor{Company}{HTML}{ccebc5}
\definecolor{University}{HTML}{ffed6f}
\definecolor{City}{HTML}{8dd3c7}
\definecolor{Tag}{HTML}{fccde5}
\definecolor{Country}{HTML}{ffffb3}

\definecolor{grey}{rgb}{0.52, 0.52, 0.51}
\definecolor{red}{rgb}{0.7, 0.11, 0.11}
\definecolor{blue}{rgb}{0.0, 0.0, 0.55}
\definecolor{green}{rgb}{0.0, 0.42, 0.24}

\usetikzlibrary{patterns}

\newcolumntype{Y}{>{\raggedright\arraybackslash}X}
\newcolumntype{R}[1]{>{\raggedleft\let\newline\\\arraybackslash\hspace{0pt}}m{#1}}
\newcolumntype{C}[1]{>{\centering\let\newline\\\arraybackslash\hspace{0pt}}m{#1}}

\newcommand{\minuscell}{\multicolumn{1}{c}{$-$}}
\newcommand{\minuscellbar}{\multicolumn{1}{c|}{$-$}}

\newcommand{\myastcellbar}{\multicolumn{1}{c|}{\myast}}

\newcommand{\memberscouncil}{Members Council\xspace}
\newcommand{\memberspolicycouncil}{Members Policy Council\xspace}

\usepackage{listings}

\usepackage{textcomp}
\definecolor{keyword}{HTML}{2771a3}
\definecolor{pattern}{HTML}{b53c2f}
\definecolor{string}{HTML}{be681c}
\definecolor{relation}{HTML}{7e4894}
\definecolor{variable}{HTML}{107762}
\definecolor{comment}{HTML}{8d9094}

\lstdefinelanguage{cypher}
{
	morekeywords={
		MATCH, OPTIONAL, WHERE, NOT, AND, OR, XOR, RETURN, DISTINCT, ORDER, BY, ASC, ASCENDING, DESC, DESCENDING, UNWIND, AS, UNION, WITH, ALL, CREATE, DELETE, DETACH, REMOVE, SET, MERGE, SET, SKIP, LIMIT, IN, CALL, CASE, WHEN,
		INDEX, DROP, UNIQUE, CONSTRAINT, EXPLAIN, PROFILE, START, FOREACH, %
		GROUP, HAVING,
	},
	sensitive=true,
	morecomment=[l]{//},
	morecomment=[s]{/*}{*/},
	morestring=[b]{"},
	literate=*{<<}{\guillemotleft{}}{1}{>>}{\guillemotright{}}{1},
}

\newcommand{\mycdots}{\cdot\!\cdot\!\cdot}
\begin{document}

\title{The Linked Data Benchmark Council (LDBC):\protect\\Driving Competition and Collaboration\protect\\ in the Graph Data Management Space}

\author{Gábor Szárnyas\inst{1}* \and
Brad Bebee\inst{2} \and
Altan Birler\inst{3} \and
Alin Deutsch\inst{4,5} \and
George Fletcher\inst{6} \and
Henry A. Gabb\inst{7} \and
Denise Gosnell\inst{2} \and
Alastair Green\inst{8} \and
Zhihui Guo\inst{9} \and
Keith W. Hare\inst{8} \and
Jan Hidders\inst{10} \and
Alexandru Iosup\inst{11} \and
Atanas Kiryakov\inst{12} \and
Tomas Kovatchev\inst{12} \and
Xinsheng Li\inst{13} \and
Leonid Libkin\inst{14} \and
Heng Lin\inst{9} \and
Xiaojian Luo\inst{15} \and
Arnau Prat-Pérez\inst{16} \and
David P\"{u}roja\inst{1} \and
Shipeng Qi\inst{9} \and
Oskar van Rest\inst{17} \and
Benjamin A. Steer\inst{18} \and
Dávid Szakállas\inst{19} \and
Bing Tong\inst{20} \and
Jack Waudby\inst{21} \and
Mingxi Wu\inst{5} \and
Bin Yang\inst{13} \and
Wenyuan Yu\inst{15} \and
Chen Zhang\inst{20} \and
Jason Zhang\inst{13} \and
Yan Zhou\inst{20} \and \\
Peter Boncz\inst{1}
}
\institute{
    \textsuperscript{1}~CWI, the Netherlands,
    \textsuperscript{2}~Amazon Web Services,
    \textsuperscript{3}~Technische Universität München, Germany,
    \textsuperscript{4}~UC San Diego,
    \textsuperscript{5}~TigerGraph,
    \textsuperscript{6}~TU Eindhoven,
    \textsuperscript{7}~Intel Corporation,
    \textsuperscript{8}~JCC Consulting,
    \textsuperscript{9}~Ant Group,
    \textsuperscript{10}~Birkbeck, University of London,
    \textsuperscript{11}~VU Amsterdam, the Netherlands,
    \textsuperscript{12}~Ontotext AD,
    \textsuperscript{13}~Ultipa,
    \textsuperscript{14}~University of Edinburgh; RelationalAI,
    \textsuperscript{15}~Alibaba Damo Academy,
    \textsuperscript{16}~\emph{work done while at UPC Barcelona and Sparsity},    
    \textsuperscript{17}~Oracle, USA,
    \textsuperscript{18}~Pometry Ltd.,
    \textsuperscript{19}~\emph{individual contributor},
    \textsuperscript{20}~CreateLink,
    \textsuperscript{21}~Newcastle University, School of Computing
    \\
    * Corresponding author, \email{gabor.szarnyas@ldbcouncil.org}
}

\maketitle

\begin{abstract}
Graph data management is instrumental for several use cases such as recommendation, root cause analysis, financial fraud detection, and enterprise knowledge representation. 
Efficiently supporting these use cases yields a number of unique requirements, including the need for a concise query language and graph-aware query optimization techniques.
The goal of the Linked Data Benchmark Council (LDBC) is to design a set of standard benchmarks that capture representative categories of graph data management problems, making the performance of systems comparable and facilitating competition among vendors.
LDBC also conducts research on graph schemas and graph query languages.
This paper introduces the LDBC organization and its work over the last decade.

\end{abstract}

\section{Introduction}
\label{sec:introduction}

\paragraph{The graph data management space}
The category of data management software with \emph{graph features} has grown steadily in the last 15~years~\cite{DBLP:journals/cacm/SakrBVIAAAABBDV21}.
This includes
graph databases~\cite{DBLP:journals/corr/abs-1910-09017},
relational DBMSs with graph extensions~\cite{DBLP:journals/debu/ZhaoY17},
graph analytics libraries~\cite{DBLP:journals/tkde/KalavriVH18},
and
graph streaming systems~\cite{DBLP:journals/corr/abs-1912-12740}.
While graph data management systems are already popular for several use cases,
such as financial fraud detection, recommendation, and data integration~\cite{DBLP:journals/vldb/SahuMSLO20}, they did not yet reach mass adoption.
We believe the two key obstacles to this are:
(1)~the lack of standardized query languages and APIs~\cite{DBLP:journals/vldb/SahuMSLO20},
(2)~limited and/or unpredictable performance in systems~\cite{DBLP:journals/cacm/SakrBVIAAAABBDV21}.
LDBC makes significant efforts to address these problems.

\paragraph{Query languages}
The adoption of graph processing systems, particularly those supporting the \emph{property graph data model}, is considerably hindered by the lack of a standard query language~\cite{DBLP:journals/vldb/SahuMSLO20}.
Currently, systems use several different query languages, including Cypher, Gremlin, GSQL, PGQL, and DQL,
which causes (potential) customers concern over lack of portability.
Starting in 2017, a concentrated effort was launched to create standard query languages.
The \SQLPGQ (Property Graph Queries) extension was released as part of SQL:2023 and
the standalone GQL (Graph Query Language) is scheduled to be released in 2024.
Both of these languages have been influenced by LDBC's \gcore design language
and LDBC has been involved in their design via its liaison with ISO.

\paragraph{Performance challenges}
Graph processing problems, including
graph pattern matching~\cite{DBLP:conf/sigmod/MhedhbiLKWS21},
graph traversal (navigation)~\cite{DBLP:journals/csur/AnglesABHRV17}, and
graph mining~\cite{DBLP:journals/pvldb/BestaVSSKGBJHLT21},
have irregular memory access patterns and provide little spatial locality or opportunities for data reuse~\cite{DBLP:conf/sc/ChecconiPWLCS12}.
Contemporary CPUs are ill-suited to handle these workloads, leading to performance problems~\cite{DBLP:journals/debu/ShaoLWX17}.
Moreover, while there were attempts to harness modern hardware such as
GPUs~\cite{DBLP:journals/csur/ShiZZJHLH18}
and
FPGAs~\cite{DBLP:journals/corr/abs-1903-06697},
these only proved beneficial for narrow domains and did not generalize to a wider set of use cases.

\paragraph{The importance of benchmarks}
To expedite the speed of progress in graph data management systems,
a group of industry and academic organizations founded the Linked Data Benchmark Council (LDBC).
LDBC is an independent benchmarking organization,
which defines standard benchmarks to make graph query performance measurable
and thus facilitate competition between vendors.
In this sense, LDBC aims to fulfill a role similar to the Transaction Processing Performance Council (TPC), which defined a number of influential benchmarks.
LDBC uses TPC's design and auditing processes as inspiration for its operations.

\paragraph{LDBC benchmarks}
LDBC has six main benchmark workloads covering different aspects of graph processing with different transactional characteristics, set of CRUD operations, and data distributions.
With the exception of Graphalytics, a leaderboard-style benchmark, all benchmarks define stringent \emph{auditing processes} for ensuring that implementations are faithful to the specification and the derived results are reproducible.
As of August 2023, LDBC published 45~audited results.

\paragraph{Paper structure}
This paper is structured as follows.
\autoref{sec:ldbc-organization} gives an overview of the LDBC organization, including its history and structure.
\autoref{sec:benchmark-task-forces} presents LDBC's benchmarks,
\autoref{sec:benchmark-processes} describes the benchmark creation and auditing processes,
and
\autoref{sec:lessons-learnt} summarizes our benchmark design experiences.
\autoref{sec:working-groups} introduces LDBC's working groups
and \autoref{sec:conclusion} outlines future directions.

\section{The LDBC organization}
\label{sec:ldbc-organization}

\subsection{History of the organization}
\label{sec:history}

\paragraph{Research project (2012--2015)}
LDBC started as a European Union-funded research project by the same name%
\footnote{FP7-ICT grant ID~317548, \url{https://cordis.europa.eu/project/id/317548}}
with the participation of 4~academic and 4~industry partners~\cite{DBLP:journals/dbsk/BonczFGL013,DBLP:journals/sigmod/AnglesBLF0ENMKT14}.
The project was coordinated by Josep Larriba Pey from Universitat Politècnica de Catalunya and Peter Boncz (CWI \& VU Amsterdam).
The consortium designed the Social Network Benchmark suite (SNB, \autoref{sec:snb}), releasing its first workload, SNB Interactive~\cite{DBLP:conf/sigmod/ErlingALCGPPB15}, and the Semantic Publishing Benchmark~\cite{DBLP:conf/semweb/KotsevMPEFK16} (SPB, \autoref{sec:spb}).
The non-profit company ``Linked Data Benchmark Council'' was established and registered in the UK.%
\footnote{\url{https://find-and-update.company-information.service.gov.uk/company/08716467}}

\paragraph{Sustained research efforts (2016--2018)}
After the EU project concluded, research efforts continued with the participation of industry partners and resulted in \gcore~\cite{DBLP:conf/sigmod/AnglesABBFGLPPS18},
a declarative language designed to formulate composable graph queries.
The Graphalytics benchmark (\autoref{sec:graphalytics})~\cite{DBLP:journals/pvldb/IosupHNHPMCCSAT16} was released,
and a draft version of the SNB Business Intelligence workload was published~\cite{DBLP:conf/grades/SzarnyasPAMPKEB18}.

\paragraph{Expansion and auditing ramp-up (2019--2022)}
LDBC's membership increased from 7~organizations in 2019 to 22~organizations in 2022 (\autoref{sec:organizational-structure}).
While in the previous phase LDBC was economically supported by CWI and Sparsity, one of the organizational improvements realized by Alastair Green was to get LDBC a bank account, to start collecting membership fees.\footnote{This seemingly trivial matter posed a practical hurdle for an organization with many directors (one per member at the time) located in different parts of the world.}
New working groups were established to research property graph schemas and query language semantics (\autoref{sec:working-groups}).
The SNB Business Intelligence workload was completed~\cite{DBLP:journals/pvldb/SzarnyasWSSBWZB22} (\autoref{sec:snb-bi-workload}).
A new Task Force was set up to design the FinBench~\cite{FinBenchWorkCharter} (\autoref{sec:finbench}).
LDBC's benchmark adoption process (\autoref{sec:defining-new-benchmarks}) and auditing processes crystallized (\autoref{sec:auditing-process}) with multiple audits occurring per year.
The term ``LDBC benchmark result'' was trademarked (\autoref{sec:trademark}).

\paragraph{Restructuring and new benchmarks (2023--)}
In early 2023, the organization was restructured to simplify governance (\autoref{sec:governance}).
The benchmark Task Forces released the initial version of the FinBench,
updated the SNB Interactive workload (\autoref{sec:snb-interactive-workload-v2}),
and organized a Graphalytics competition (\autoref{sec:graphalytics-competition}).

\subsection{Organizational structure and operations}
\label{sec:organizational-structure}

\paragraph{Historical structure (2013--2022)}
Until 2023, LDBC member organizations could appoint a director to the \emph{Board of Directors}, which at its peak consisted of 20\texttt{+} members,
an unusual and unwieldy structure for a small non-profit company.

\begin{figure}[htb]
    \vspace{-2.5ex}
    \centering
    \includegraphics[scale=\yedscale]{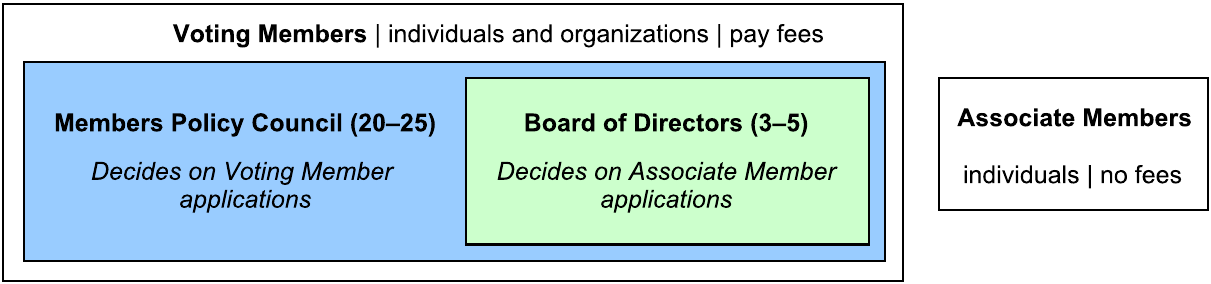}
    \caption{Organizational structure of the LDBC from May 2023.}
    \label{fig:organizational-structure}
    \vspace{-2.5ex}
\end{figure}

\paragraph{Current structure (2023--)}
\label{sec:governance}
To simplify its governance, LDBC was restructured in 2023, resulting in new
articles of association~\cite{LdbcArticlesOfAssociation}
and
updated byelaws~\cite{LdbcByelaws}.
The new structure (\autoref{fig:organizational-structure}) has \emph{Voting Members}, who contribute via the \emph{\memberspolicycouncil}%
\footnote{The \emph{\memberspolicycouncil} is called the \emph{\memberscouncil} in official documents.}
and \emph{Associate Members}, who pay no fees (and have no vote) but contribute to the day-to-day work of LDBC.
There is a new, smaller \emph{Board of Directors} (3--5 members), who are also part of the \memberspolicycouncil.

\paragraph{Membership}
As of August 2023,
LDBC has 24~member organizations, including database, hardware, and cloud computing vendors, and academic institutes.
There are 3~sponsor companies\footnote{%
Ant Group,
Beijing Volcano Engine Technology Co.,
and
Oracle Labs%
},
18~member companies\footnote{%
Amazon,
Alibaba Damo Academy,
ArangoDB,
Beijing Haizhi Xingtu Company,
CreateLink,
Fabarta,
Intel,
JCC Consulting,
Katana Graph,
Memgraph,
Neo4j,
Ontotext,
Pometry,
RelationalAI,
Sparsity Technologies,
TigerGraph,
Ultipa,
and
vesoft%
},
and
3~non-commercial institutions\footnote{
FORTH;
Birkbeck, University of London;
and
Zhejiang Lab
}.
LDBC has
3~individual voting members%
\footnote{Peter Boncz, Alexandru Iosup, and Gábor Szárnyas; all co-authors of this paper}
and
60\texttt{+}~associate members.

\paragraph{Intellectual property rights}
The intellectual property rights policies of LDBC cover assets created by members while participating in LDBC activities, including software components, written specifications, discussion proposals, academic papers, \etc~\cite{LdbcByelaws}. Software contributions are licensed under a licence substantively identical to the Apache Software Licence 2.0; copyright for documentary or pictorial contributions are licensed to LDBC with a right to sublicence. Typically LDBC publishes contributions either to its members or to external standards groups like ISO, or generally to the public under the Creative Commons CC-BY licence.
Participants in activities of LDBC are also required to comply with the LDBC Patent Rules, which include disclosing patents that may be infringed by the implementation of an LDBC benchmark specification, or by an external standard that incorporates contributions from LDBC.

\paragraph{Teams}
LDBC's members form teams that work on specific aspects of graph processing.
\emph{Task forces} design, implement, and maintain benchmarks (\autoref{sec:benchmark-task-forces}).
\emph{Working groups} conduct research on graph query languages and graph schema (\autoref{sec:working-groups}).
The establishment of these teams is initiated by the creation of a \emph{work charter} and is voted on by the \memberspolicycouncil.

\paragraph{Finances}
LDBC's sources of revenue are its membership and auditing fees.
As of 2023, membership costs
\numprint{1100} GBP for non-commercial institutes,
\numprint{2200} GBP for commercial companies, and
\numprint{8800} GBP for sponsor members.
A fee of \numprint{2000} GBP is applied to audits %
commissioned by non-sponsor members.
Individual associate membership is free.
LDBC uses its funds to pay for cloud compute, storage, and collaboration services in addition to usual company overheads.

\subsection{Liaison with ISO on standard query languages (GQL, SQL/PGQ)}
\label{sec:liaison}
\label{sec:standard-query-languages}

In 2017,  the international standards committee responsible for the SQL database language standard (ISO\slash IEC\slash JTC~1\slash SC~32\slash WG~3 ``Database Languages'')%
\footnote{\url{https://www.iso.org/organization/6720817.html}},
established a Category~C Liaison relationship with LDBC.
This liaison allows WG~3 to share its working documents, draft specifications, and draft digital artifacts with LDBC participants.
This access gives LDBC members early visibility into standards development efforts.

Peter Boncz, Alastair Green, and Jan Hidders are LDBC's liaison participants on the official ISO roster for WG~3, and can participate in any WG~3 meeting.
This liaison relationship has helped to make the SC~32/WG~3 work more visible to organizations participating in LDBC and give the LDBC work more visibility in the Database Language standards process. 

WG~3 has been working on two projects that are of interest to LDBC members:
\begin{itemize}
    \item ISO/IEC~9075-16 Information technology --- Database languages SQL --- Part~16: Property Graph Queries (\SQLPGQ)
    \item ISO/IEC~39075 Information technology --- Database languages --- GQL
\end{itemize}

\SQLPGQ is completed and was published by ISO at the end of May 2023.
GQL is currently undergoing a Draft International Standard (DIS) ballot that started on 2023-05-23 and
ends on 2023-08-15.
WG~3 aims to resolve any issues identified during the DIS ballot and to have the GQL standard ready for publication in early 2024.

Within the database language standards committees, there is ongoing work to expand the Graph Pattern Matching (GPM) language~\cite{DBLP:conf/sigmod/DeutschFGHLLLMM22} in areas such as cheapest path queries. This GPM work will be integrated into the next editions of both \SQLPGQ and GQL.

The LDBC Extended Schema (LEX) working group~\cite{LDBC:WC:WC-2022-02} (\autoref{sec:lex}) aims to propose expanded schema capabilities in a future edition of the GQL standard.

\subsection{Technical User Community (TUC) meetings}
\label{sec:tuc}

\begin{table}[htb]
    \vspace{-2.5ex}
    \centering
    \caption{LDBC Technical User Community meetings between 2018 and 2023.}
    \label{tab:tuc}
    \setlength\tabcolsep{3.0pt}
    \begin{tabular}{@{}rclllr@{}}
        \toprule
        \thead{\texttt{\#}} & \thead{Year} & \thead{Date}    & \thead{Location}           & \thead{Format} & \thead{Program} \\
        \midrule
        16                  & 2023         & June 23--24     & Seattle, WA                & hybrid         & 31~talks        \\
        15                  & 2022         & June 17--18     & Philadelphia, PA           & hybrid         & 26~talks        \\
        14                  & 2021         & August 16       & Copenhagen, Denmark        & hybrid         & 19~talks        \\
        13                  & 2020         & June 30--July 1 & \emph{online}              & online         & 4~sessions      \\
        12                  & 2019         & June 5          & Amsterdam, the Netherlands & in-person      & 13~talks        \\
        11                  & 2018         & June 8          & Austin, TX                 & in-person      & 11~talks        \\
        \bottomrule
    \end{tabular}
    \vspace{-2.5ex}
\end{table}

Since 2012, LDBC organizes Technical User Community (TUC) meetings.%
\footnote{\url{https://ldbcouncil.org/tags/tuc-meeting/}}
These are 1--2~day informal workshops,
where LDBC's leaders, task forces, and working groups report on their progress.
Additionally, member companies give updates of their products,
researchers in the graph space discuss their latest results,
and users of graph data management systems present their use cases.
The meetings provide an opportunity for members to contribute to LDBC's \emph{choke points} (\autoref{sec:benchmark-terminology}), and to influence the future direction of LDBC.
In recent years, the TUC meetings have been steadily gaining popularity (\autoref{tab:tuc}).

\section{Benchmarks}
\label{sec:benchmark-task-forces}
\label{sec:choke-points}

\begin{table}[htb]
    \vspace{-2.5ex}
    \centering
    \caption{
        Key characteristics of LDBC benchmarks.
        \emph{Scale:} size of the largest data set, %
        \emph{GP lang.:} is the use of general-purpose programming languages allowed for implementations?
        \emph{Req. isol.:} required isolation level (\textsf{SI}: snapshot isolation, \textsf{RC}: read committed).
        Legend:
        \yes~yes,
        \no~no,
        \maybe~optional;
        \myast~the benchmark is under design and audits are not yet possible;
        (i)~larger sizes can be generated using the Graphalytics graph generator, but are not part of the standard benchmark.
    }
    \label{tab:benchmark-characteristics}
    \setlength\tabcolsep{3.5pt}
    \footnotesize
    \begin{tabular}{@{}l|c|l|R{1cm}|r|r|r|cc|cc|l@{}}
        \toprule
        \theadbar{Benchmark}      &
        \theadbar{Year}           &
        \theadbar{Workload}       &
        \theadbar{Scale}          &
        \theadsidebar{Min. scale} &
        \theadsidebar{\#Queries}  &
        \theadsidebar{\#Audits}   &
        \theadside{Inserts}       &
        \theadsidebar{Deletes}    &
        \theadside{GP lang.}      &
        \theadsidebar{ACID test}  &
        \theadside{Req. isol.}                                                                                                                                                    \\
        \midrule
        SNB BI                    & 2022 & analytical    & \ThirtyK                       & 30            & 20 & 4             & \yes & \yes & \no  & \maybe        & \textsf{SI} \\ %
        SNB Interactive v1        & 2015 & transactional & \OneK                          & 30            & 21 & 24            & \yes & \no  & \yes & \yes          & \textsf{RC} \\ %
        SNB Interactive v2        & \myast & transactional & \ThirtyK                       & 30            & 21 & \myastcellbar & \yes & \yes & \yes & \yes          & \textsf{SI} \\ %
        \midrule
        SPB                       & 2015 & transactional & 5                             & 1            & 12 & 17            & \yes & \yes & \no  & \no           & \textsf{RC} \\ %
        FinBench       & 2023 & transactional & 10                             & 0.1            & 40 & 0 & \yes & \yes & \yes & \yes          & \textsf{RC} \\
        Graphalytics              & 2016 & algorithms    & \numprint{320}$\information{}$ & \minuscellbar & 6  & \minuscellbar & \no  & \no & \yes  & \minuscellbar & \minuscell  \\ %
        \bottomrule
    \end{tabular}
    \vspace{-2.5ex}
\end{table}

In this section, we describe LDBC's benchmarks.
We first present LDBC's common benchmark terminology (\autoref{sec:benchmark-terminology}).
We then describe the workloads of the Social Network Benchmark suite (SNB, \autoref{sec:snb}),
followed by the Semantic Publishing Benchmark (SPB, \autoref{sec:spb}),
the FinBench (\autoref{sec:finbench}),
and
Graphalytics (\autoref{sec:graphalytics}).
The benchmarks are summarized in \autoref{tab:benchmark-characteristics}.
Systems that implement at least two LDBC benchmarks are shown in \autoref{tab:benchmark-implementations}.

\newcolumntype{Q}{m{0.58cm}}
\newcolumntype{P}{>{\centering\arraybackslash}m{0.58cm}}

\begin{table}[htb]
    \vspace{-2.5ex}
    \centering
    \caption{
        Systems with implementations for 2\texttt{+} LDBC benchmarks.
        Legend:
        \yes~full,
        \maybe~partial,
        \no~no implementation;
        (a)~audited implementation;
        (s)~the implementation was used for a standard-establishing audit.
    }
    \label{tab:benchmark-implementations}
    \setlength\tabcolsep{3.1pt}
    \footnotesize
    \begin{tabular}{@{}l|Q|Q|Q|P|P|Q|P|Q|Q|Q|Q@{}}
        \toprule
        \theadbar{Benchmark} & \theadsidebarleftdouble{CreateLink}{Galaxybase} & \theadsidebarleftdouble{Ontotext}{GraphDB} & \theadsidebarleftdouble{Graphscope-}{Flex} & \theadsidebarleftsingle{Neo4j} & \theadsidebarleftsingle{PostgreSQL} & \theadsidebarleftdouble{Sparsity}{Sparksee} & \theadsidebarleftdouble{Microsoft}{SQL Server} & \theadsidebarleftsingle{TigerGraph} & \theadsidebarleftdouble{Ant Group}{TuGraph} & \theadsidebarleftsingle{Umbra} & \theadsideleft{OpenLink}{Virtuoso} \\\midrule
        SNB BI               & \no                                       & \no                                  & \no                                 & \yes                           & \yes                                & \maybe                                & \no                                      & \yes$\stdestablishing\audited$      & \no                                   & \yes$\stdestablishing$         & \maybe                             \\
        SNB Interactive v1   & \yes$\audited$                            & \yes$\audited$                       & \yes$\audited$                      & \yes                           & \yes                                & \yes$\stdestablishing$                & \maybe                                   & \yes                                & \yes$\audited$                        & \yes                           & \yes$\stdestablishing$             \\
        SNB Interactive v2   & \no                                       & \no                                  & \no                                 & \yes                           & \yes                                & \no                                   & \yes                                     & \no                                 & \maybe                                & \yes                           & \no                                \\\midrule
        SPB                  & \no                                       & \yes$\audited$                       & \no                                 & \no                            & \no                                 & \no                                   & \no                                      & \no                                 & \no                                   & \no                            & \yes$\audited$                     \\
        FinBench             & \yes$\stdestablishing$                    & \no                                  & \no                                 & \no                            & \no                                 & \no                                   & \no                                      & \no                                 & \yes$\stdestablishing$                & \no                            & \no                                \\
        Graphalytics         & \yes                                      & \no                                  & \yes                                & \yes                           & \no                                 & \no                                   & \no                                      & \no                                 & \yes                                  & \maybe                         & \no                                \\
        \bottomrule
    \end{tabular}
    \vspace{-2.5ex}
\end{table}

\subsection{Benchmark terminology}
\label{sec:benchmark-terminology}

\paragraph{Choke points}
\label{sec:choke-points}
LDBC's benchmark design process uses \emph{choke points}~\cite{DBLP:conf/tpctc/BonczNE13}, \ie well-chosen technical difficulties that are challenging for the present generation of data processing systems and whose optimization likely results in significant overall performance improvements.
The choke points are identified by expert data systems architects and also subject to feedback received at the Technical User Community meetings (\autoref{sec:tuc}).
LDBC workloads are designed to cover the set of choke points triggered by a given workload category.

\paragraph{Auditing}
\label{sec:auditing}
Similarly to TPC's \emph{Enterprise Class benchmarks}~\cite{DBLP:conf/tpctc/Poess22}, most of LDBC's benchmarks must undergo an auditing process conducted by a certified auditor before they can be published as official results.
The auditors check compliance with the specification,
run the benchmark independently
and present their findings in a \emph{full disclosure report} (FDR).
The FDR documents the benchmark setup and the derived results in detail, typically spanning over 20--50 pages.
Additionally, audited benchmark results are accompanied by a \emph{supplementary package}, which includes the benchmark implementation and the binary of the system-under-test (SUT), ensuring that the results are reproducible.

\paragraph{Scale factors}
\label{sec:scale-factors}
LDBC's benchmark suites include data generators that produce synthetic data sets of increasing sizes.
Each data set is characterized by its \emph{scale factor} (SF) which corresponds to the data set's disk usage
when serialized in CSV (comma-separated values) format, measured in GiB.

\paragraph{ACID compliance}
\label{sec:acid}
Several of LDBC's benchmarks require the SUT to comply with ACID properties. %
An important aspect of this is \emph{durability}:
the SUT must be able to recover from a crash or power outage without losing any committed data.%
\footnote{Unlike TPC, LDBC does not require systems to tolerate hardware failures.}
For the SNB workloads and FinBench, the \emph{isolation} properties are tested with an ACID test suite.

\subsection{Social Network Benchmark (SNB) suite}
\label{sec:snb}

The Social Network Benchmark suite pioneered a number of techniques used in LDBC benchmarks:
choke point-driven design~\cite{DBLP:conf/tpctc/BonczNE13},
scalable correlated dynamic graph generation~\cite{DBLP:conf/tpctc/PhamBE12,DBLP:conf/sigmod/WaudbySPS20},
and
parameter curation for stable query runtimes~\cite{DBLP:conf/tpctc/GubichevB14}.
The detailed specification of the SNB workloads is available~at~\cite{DBLP:journals/corr/abs-2001-02299}.

\paragraph{SNB data generators}
The first version of the SNB data generator was implemented in Hadoop and only supported insert operations~\cite{DBLP:conf/tpctc/PhamBE12}.
In 2020, it was ported to Spark for improved scalability,%
\footnote{\url{https://github.com/ldbc/ldbc_snb_datagen_spark}}
and was extended with support for producing deep (cascading) delete operations~\cite{DBLP:conf/sigmod/WaudbySPS20}.
To the best of our knowledge, its ability to generate a scalable graph where structure and values correlate, with flashmob-style spikes and  %
deep delete operations are features unique to the SNB data generator~\cite{DBLP:journals/csur/BonifatiHPS20}.

\subsection*{SNB Interactive workload v1}
\label{sec:snb-interactive-workload-v1}

The SNB \interactivevone workload was published in 2015~\cite{DBLP:conf/sigmod/ErlingALCGPPB15}.
It is a transactional benchmark that targets OLTP systems with graph features (\eg path-finding).
The workload consists of three types of operations:
14~complex read queries,
7~short read queries, and
8~inserts.
The workload has a balanced mix of operations with approximately 8\% complex reads, 72\% short reads, and 20\% inserts.

\begin{figure}[htb]
    \vspace{-2.5ex}
    \centering
    \includegraphics[width=\textwidth]{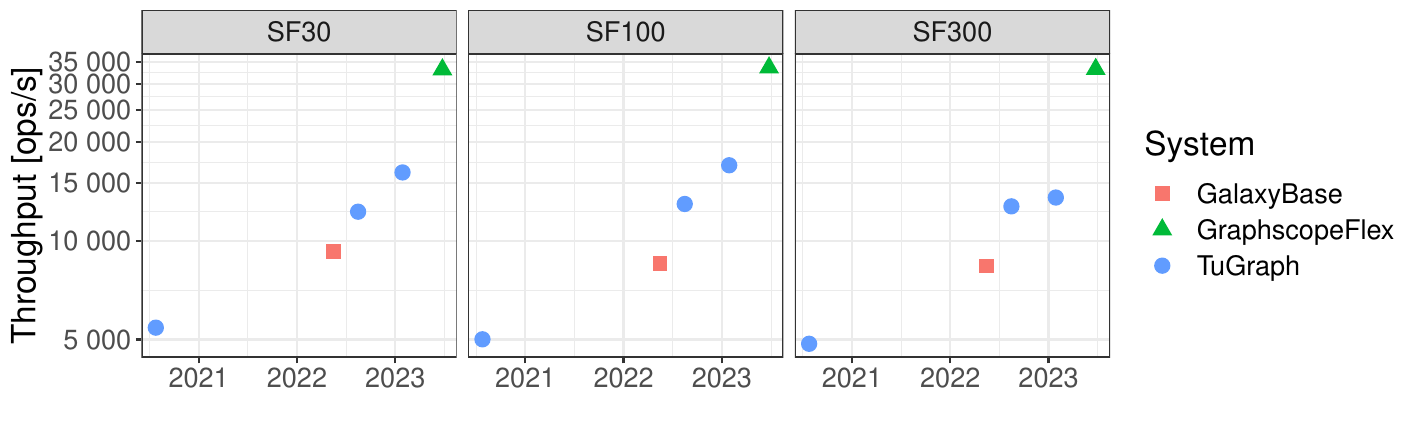}
    \vspace{-2.5ex}
    \caption{
        Throughput of the audited results %
        over time for the \snbinteractivevone workload for scale factors 30, 100, 300 for implementations using general-purpose programming languages (C\texttt{++} and Java), obtained between 2021 and 2023.}
    \label{fig:interactive-audited-results}
    \vspace{-2.5ex}
\end{figure}

\snbinteractivevone has been influential in the graph data management space:
as of June 2023, 24~audited results were published using this workload.%
\footnote{\url{https://ldbcouncil.org/benchmarks/snb-interactive/}}
\autoref{fig:interactive-audited-results} shows the top 15~audited results published between 2021 and 2023, which demonstrate a performance increase of more than $6\times$ over this period.

\subsection*{SNB Business Intelligence workload}
\label{sec:snb-bi-workload}

The SNB Business Intelligence (BI) workload~\cite{DBLP:journals/pvldb/SzarnyasWSSBWZB22} captures an OLAP scenario with heavy-hitting analytical queries that touch on large portions of the graph (\eg aggregating all \Messages created within a 100-day period or exploring the neighbourhoods of the \Persons living in China).
The queries include multi-source \emph{cheapest path-finding} (also known as \emph{weighted shortest paths}), cyclic graph patterns, and have correlated and anti-correlated variants.
The workload uses an improved data generator which produces fully dynamic graphs with insert and delete operations~\cite{DBLP:conf/sigmod/WaudbySPS20}, and scales to SF\ThirtyK.
The BI workload targets both DBMSs and data analytical systems such as Spark.
To this end, updates can be applied in two modes:
in \emph{concurrent read--write mode}, reads and writes are executed concurrently,
while in \emph{disjoint read--write mode}, alternating blocks of reads and writes (accounting for one day's worth of updates) are executed. 
The BI workload was officially approved in November 2022.
Since then, it has accumulated 4~audited results, including one for SF\TenK.
\footnote{\url{https://ldbcouncil.org/benchmarks/snb-bi/}}

\subsection*{SNB Interactive workload v2}
\label{sec:snb-interactive-workload-v2}

In 2022, the SNB task force initiated the renewal of the \interactivevone workload.
The \interactivevtwo workload adopted several features from the BI workload:
support larger scale factors and delete operations,
and coverage of the \emph{cheapest path-finding} algorithm.
Moreover, the updated workload introduces a new temporal parameter curation algorithm to ensure stable runtimes for path queries evaluated on a fully dynamic graph,
and features a complete refactoring of the driver along with several usability improvements.
The balance of the read and insert queries are similar to \interactivevone workload with only 0.2\% delete operations added -- motivated by the fact that deletes are rare compared to inserts in most real-life systems~\cite{DBLP:conf/cscw/AlmuhimediWLSA13}.
As of June 2023, the workload is in draft phase with four complete implementations already available (\autoref{tab:benchmark-implementations}).

\subsection{Semantic Publishing Benchmark (SPB)}
\label{sec:spb}

The Semantic Publishing Benchmark (SPB) targets RDF database engines and is inspired by the BBC's Dynamic Semantic Publishing approach~\cite{DBLP:conf/semweb/KotsevMPEFK16}. The scenario behind the benchmark considers a media that deals with a large volume of streaming content, namely articles and other ``creative works''. This content is enriched with metadata that describes it and links it to reference knowledge -- taxonomies and databases that include relevant concepts, entities, and factual information. This metadata allows publishers to efficiently retrieve relevant content and increase engagement.

The main interactions with the repository are
(i)~updates, which add new creative works or alter the repository, and
(ii)~aggregation queries, which retrieve content. The engine should handle updates executed concurrently with a massive amount of aggregation queries. It must be able to deal with graph pattern matching, reasoning, geospatial constraints, and full-text search. The SPB includes ontologies, a data generator, query patterns, as well as test and benchmark drivers. 
The detailed specification of SPB version~2.0 is available at~\cite{ldbc-spb-specification}.

\subsection{Graphalytics}
\label{sec:graphalytics}

The LDBC Graphalytics benchmark targets systems that evaluate graph algorithms on static graphs.
While the scope of Graphalytics is similar to the influential GAP Benchmark Suite~\cite{DBLP:journals/corr/BeamerAP15}, Graphalytics has slightly different algorithms, more data sets (38 \vs 5), and enforces determinism in its algorithms.

\paragraph{Algorithms}
The selection of graph algorithms for Graphalytics was motivated by input received during the TUC meetings (\autoref{sec:tuc}) and a literature survey of over 100~academic papers~\cite{DBLP:conf/ipps/GuoBVIMW14}.
The benchmark includes the following six algorithms:
breadth-first search (BFS),
community detection using label propagation (CDLP),
local clustering coefficient (LCC),
PageRank (PR),
single-source shortest paths (SSSP),
and
weakly connected components (WCC).
Algorithms were adjusted if necessary such that they are \emph{deterministic:}
BFS returns the \emph{distance} (level) for each node in the traversal (instead of picking one of the parent nodes),
while
the tie-breaking strategy employed in the CDLP algorithm always picks the smallest label (instead of picking randomly).

\paragraph{Data set}
Graphalytics uses untyped and unattributed graphs.
The data set of the benchmark includes both directed and undirected graphs.
Some graphs have edge weights, which are used exclusively by the SSSP algorithm.

\paragraph{Graphalytics competition}
\label{sec:graphalytics-competition}
Presently, it is not possible to commission audits for Graphalytics.
Instead, LDBC organizes competitions with leaderboards,
in the style of the Top500%
\footnote{\url{https://www.top500.org/}} and Graph500%
\footnote{\url{https://graph500.org/}}
competitions of the high-performance computing community.
Solutions compete for both performance (execution time) and scalability.
Implementers are also required to report system prices following the TPC Pricing Specification~\cite{tpc-pricing}.

\subsection{FinBench}
\label{sec:finbench}

The FinBench (short for ``Financial Benchmark'') is a benchmark, new in 2023, specifically designed to evaluate the performance of graph database systems in financial scenarios, such as anti-fraud and risk control, by employing financial data patterns and query patterns. This collective effort led by Ant Group adopts its rich experience in financial services, making it more applicable to graph users in the financial industry than previous LDBC benchmarks. Sharing some framework similarities with SNB, it distinguishes itself through differences in datasets and workloads.

\paragraph{Data pattern}
Financial graphs are distinguished from the social network in SNB by having hub vertices with higher degrees and allowing edge multiplicity.
Financial graphs are asymmetric directed graphs causing more unbalanced load and less spatial locality. The maximum degrees of hub vertices are in the magnitude of thousands in the social network generated by the SNB data generator, while the degree of hub vertices in the financial graphs generated by the FinBench data generator%
\footnote{\url{https://github.com/ldbc/ldbc_finbench_datagen}}
may scale up to millions in large data scales. The higher degree of hub vertices poses new challenges to the performance of systems. The edge multiplicity means multiple edges of the same type can exist between the same source vertex and destination vertex. This requires support in the system storage and provides optimization opportunities for filtering during traversal.

\paragraph{Transaction workload}
The \emph{transaction workload} is the first workload included in the initial version of FinBench targeting OLTP data management systems, as the SNB Interactive workload~\cite{DBLP:conf/sigmod/ErlingALCGPPB15} does.
The key features of transaction workload include read-write query, time-window filtering, and recursive path filtering. The read-write query is a new query type, which wraps a complex query in a transaction.
The write query to execute is determined by the result of the complex read.
Time-window filtering is a pattern when financial businesses focus on the data, especially the edges, in a specific time range. Recursive path filtering is a pattern used by financial businesses to find fund traces in the graph.
A fund trace may match the filter that the timestamp increases and the amount decreases of the edges sourcing from the origin vertex hop-by-hop.
These typical features bring new challenges to the performance of systems.
The transaction workload has 12~complex read, 6~simple read, 19~write, and 3~read-write queries.

The initial version of FinBench %
was collaboratively developed by nine leading graph system vendors. It underwent cross-validation on three systems: TuGraph, Galaxybase, and UltipaGraph. For more detailed information, please refer to the repositories on GitHub: FinBench specification%
\footnote{\url{https://github.com/ldbc/ldbc_finbench_docs}}, FinBench driver%
\footnote{\url{https://github.com/ldbc/ldbc_finbench_driver}}, reference implementation of the FinBench transaction workload%
\footnote{\url{https://github.com/ldbc/ldbc_finbench_transaction_impls}}, FinBench ACID suite%
\footnote{\url{https://github.com/ldbc/ldbc_finbench_acid}}.

\section{Benchmark processes}
\label{sec:benchmark-processes}

\subsection{Defining new LDBC benchmarks}
\label{sec:defining-new-benchmarks}

LDBC has a strict process for proposing new benchmarks.
Based on our experience, the initial benchmark completion (phases~1 to~3) takes at least 2~years, followed by adoption and maintenance (phases~4 and~5), which may span 5\texttt{+} years.

\paragraph{Phase 1: Benchmark proposal}
The benchmark proposer shall create a draft proposal.
The benchmark must be motivated by real-world use cases and target a category of data processing systems that tackle some aspect of graph processing.
The designers must reason why the benchmark is significantly different from existing LDBC benchmarks
by
identifying new performance challenges and formulating their choke points,
providing data with unique characteristics (\eg distribution, frequency/type of updates), \etc
The benchmark draft shall be presented the benchmark to the \memberspolicycouncil to gather feedback.

\paragraph{Phase 2: Collaboration setup}
The proposer shall gather agreements from 2\texttt{+} member companies who are willing to contribute to the benchmark specification and create reference implementations.
They shall create a work charter for the benchmark task force (\eg~\cite{FinBenchWorkCharter,SnbWorkCharter}), which includes the list the members interested in working on the benchmark.
This shall be presented to the \memberspolicycouncil, which votes \emph{on the establishment of a new benchmark task force}.

\paragraph{Phase 3: Detailed benchmark design}
The task force shall create the detailed benchmark specification, implement the data generator and the benchmark driver.
The creation of at least 2\texttt{+} complete reference implementations %
is required.
The task force shall ensure that the specification contains the description of the data set, queries, and workload as well as the detailed auditing guidelines.
They shall publish the specification in an open repository %
and release the software components as open-source.
The task force shall conduct 2\texttt{+} \emph{standard-establishing audits}.
These include the complete execution of the benchmark and the creation of FDR that detail their outcomes.
The task force shall submit all resulting documents to the \memberspolicycouncil, which votes \emph{on the acceptance of the benchmark}.

\paragraph{Phase 4: Adoption and auditing}
The task force shall help adoption attempts by working closely with the benchmark's early users.
They should create training material and exam questions for auditor exams, then train and certify auditors.
Certified auditors can fulfill incoming audit requests, initially with close collaboration with the benchmark task force.

\paragraph{Phase 5: Maintenance and renewal}
The task force shall maintain the benchmark and optionally assist in further adoption and auditing attempts.
If the benchmark remains popular, the task force should consider renewing it after 5--10 years to ensure its continued relevance.

\subsection{Auditing process}
\label{sec:auditing-process}

\paragraph{Phase 1: Preparation (1--2 weeks)}
The Test Sponsor shall be an LDBC member company.
If the Test Sponsor is not an LDBC sponsor member company, it shall pay LDBC an auditing fee of \numprint{2000} GBP (as of 2023).
The Test Sponsor shall create an initial version of the supplementary package of the benchmark and send it to an LDBC-certified Auditor for a preliminary review.
The Test Sponsor shall establish the costs of its system setup using the TPC Pricing Specification~\cite{tpc-pricing}.

\paragraph{Phase 2: Audit setup (3--6 weeks)}
The Test Sponsor and the Auditor shall negotiate the timeline and pricing of the audit, and sign a contract.
The Test Sponsor shall hand over the supplementary package to the Auditor.
Continuous communication in the form of emails, online meetings, DMs, \etc between the Auditor and the Test Sponsor is recommended for status updates and clarifications.

\paragraph{Phase 3: Auditing (3--10 weeks)}
For the SNB workloads, an audit consists of the following steps:
\begin{enumerate*}[(1)]
    \item set up system,
    \item run cross-validation,
    \item perform code review,
    \item run ACID isolation tests,
    \item perform recovery tests,
    \item conduct benchmark runs on the scale factors requested by the Test Sponsor,
    \item write the FDR and the executive summary.
\end{enumerate*}
The Test Sponsor then reviews the FDR and executive summary documents.
If everything is in order, the Auditor, the Test Sponsor, and the leader of the task force sign the FDR.

\paragraph{Phase 4: Dissemination of results (1--2 weeks)}
The audit results are announced on the LDBC website, on mailing lists and on social media.

\paragraph{Timespan}
The overall time required for an audit is between 8 and 20~weeks.

\subsection{Trademark}
\label{sec:trademark}

To prevent misuse of the benchmarks, the term ``LDBC benchmark result'' is trademarked and is only allowed to be used for results that were achieved by an LDBC-certified Auditor in an official audit.
That said, LDBC encourages use (including derived use) of its benchmarks provided that users comply with the \emph{fair use policies},
described in LDBC's Byelaws~\cite{LdbcByelaws}.

\section{Benchmarking lessons learnt}
\label{sec:lessons-learnt}

This section captures our key lessons learnt with designing and maintaining benchmarks, and conducting audits with them.

\paragraph{High development costs}
We found that creating domain-specific application-level benchmarks is a big undertaking.
Realistic benchmarks are bound to be complex as they require a (somewhat) realistic scalable data generator and a high-performance driver with reference implementations.
To make matters more complicated, the graph domain has highly skewed and correlated data sets, causing queries to be sensitive to parameter selection~\cite{DBLP:conf/tpctc/GubichevB14}.
Path-finding queries on fully dynamic graphs are particularly susceptible to this, necessitating expensive parameter generation steps~\cite{DBLP:journals/pvldb/SzarnyasWSSBWZB22}.
Creating reference implementations is also labour-intensive due to the lack of a standard query language -- fortunately, this is expected to improve with the introduction of \SQLPGQ and GQL.

\paragraph{Data set availability is important}
We found that users prefer downloading pre-generated data sets from an official repository instead of generating them using the data generators.
To help adoption, it is best to make the data sets available in multiple serialization formats (different layouts, datetime formats, \etc) for all scale factors.
However, this requires tens of terabytes of storage, leading to high storage costs.
Moreover, we transferring large data sets stored at public cloud providers can be prohibitively expensive due to high egress fees
(\ie fees paid for transferring data out of the cloud).
To work around this problem, we use storage services which do not have egress fees.
For long-term archiving, we store our data in the SURF Data Repository,\footnote{\url{https://github.com/ldbc/data-sets-surf-repository}} which is operated by the Dutch national HPC support center, and offers tape-based storage.
For short-term data distribution, we use Cloudflare's R2 service.

\paragraph{Shift to the cloud}
We observed a shift to the cloud for database management~\cite{DBLP:journals/cacm/AbadiAABBBBCCDD22}:
approximately half of LDBC's graph vendors have a cloud offering and some have cloud-native systems with no on-premise solutions.
The use of the cloud is also popular for running audits (and is encouraged by LDBC for easier reproducibility):
35~out of 49~audits conducted so far were executed in the cloud.

\paragraph{Finding problems beyond peformance}
While the main goal of a benchmark implementation is to measure performance and to identify bottlenecks, implementing a full workload often leads to the discovery of other issues.
Namely, we have found several issues such as
insufficient query language features,
correctness bugs,
concurrency issues on different CPU architectures,
crashes on large data sets,
durability errors,
parameter handling errors,
issues with datetime and string handling,
and
deadlocks caused by concurrent transactions.
The availability of public benchmark data sets made these errors easy to reproduce and reason about, leading to significant improvements in the systems-under-test.

\section{Working groups}
\label{sec:working-groups}

LDBC's working groups conduct research on areas related to graph query languages,
including formalization of (sub)languages and exploring possibilities for defining graph schemas.

\subsection{Graph Query Languages working group}
\label{sec:gcore}
The \emph{Graph Query Languages} working group created the \gcore design language~\cite{DBLP:conf/sigmod/AnglesABBFGLPPS18}, which treats paths as first-class citizens and supports the composability of graph queries.
While the working group ceased to exist after the publication of \gcore, LDBC has a liaison with ISO (\autoref{sec:liaison}) that facilitates continued collaboration on query language standards.

\subsection{Formal Semantics Working Group (FSWG)}
\label{sec:fswg}

The \emph{Formal Semantics Working Group} (FSWG) gives formal treatment to standard graph query languages to prevent ambiguous interpretations.
To this end, it formalized the Graph Pattern Matching (GPM) language of GQL and \SQLPGQ~\cite{DBLP:conf/sigmod/DeutschFGHLLLMM22}.
The group also produced a formal summary of the GQL language~\cite{DBLP:conf/icdt/FrancisGGLMMMPR23} and created a pattern calculus for property graphs~\cite{DBLP:conf/pods/FrancisGGLMMMPR23} that serves as a theoretical basis of GPM.

\subsection{Property Graph Schema Working Group (PGSWG)}
\label{sec:pgswg}

Initial graph database systems were schemaless, which hindered their adoption in several enterprise domains.
The \emph{Property Graph Schema Working Group} (PGSWG) investigates the problem of defining schemas for the property graph data model~\cite{DBLP:journals/corr/abs-2211-10962}.
The group also identified ways to define keys in property graphs~\cite{DBLP:conf/sigmod/AnglesBDFHHLLLM21}.

\subsection{LDBC Extended GQL Schema (LEX)}
\label{sec:lex}

The \emph{LDBC Extended GQL Schema} (LEX) working group was established in 2022 with an initial membership of 10~organizations and approximately 20~individuals~\cite{LDBC:WC:WC-2022-02}.
The group aims to propose the addition of a future extended schema definition language to the GQL standard,
which allows for more elaborate (and therefore more restrictive) constraints on the permitted values of GQL property graphs than can be imposed by graph types as defined in the GQL DIS.
These additional constraints aim to establish parity with the constraints available in SQL schema, to incorporate features described in PG-Schema~\cite{DBLP:journals/corr/abs-2211-10962} and to support performant processing of incremental transactional updates of a graph database.

\section{Conclusion and future outlook}
\label{sec:conclusion}

In this paper, we summarized LDBC's history, organization structure, community management; as well as its benchmarks, working groups, and processes.
At the time of writing (June 2023), LDBC has a healthy benchmark ecosystem, which is actively maintained and renewed.
Our benchmarks are used by a number of database vendors for both internal benchmarking as well as impartial comparisons via audits, which are now performed routinely.

In the future, we plan to further improve our benchmarks, including support for larger scale factors (up to SF\HundredK for SNB).
We also plan to investigate the impact of incorporating long-running transactions in our transactional workloads.
Finally, we are interested in creating benchmarks for new areas of graph data management such as streaming and machine learning on graphs.

\bibliographystyle{abbrv}
\bibliography{ms}

\end{document}